\documentclass[twocolumn,preprintnumbers,secnumarabic,amsmath,amssymb,nofootinbib,floatfix]{revtex4}
\usepackage{varioref,exscale,latexsym,amsmath,amssymb}
\usepackage{graphicx}

\usepackage{graphicx}
\usepackage{dcolumn}
\usepackage{bm}

\def\beq{\begin{equation}}

\def\eeq{\end{equation}}

\def\beqa{\begin{eqnarray}}

\def\eeqa{\end{eqnarray}}

\begin{document}

\title{{\bf The emergence of a universal limiting speed }}

\medskip\
\author{Mohamed M. Anber${}^{1}$}
\email[Email: ]{manber@physics.utoronto.ca}
\author{ John F. Donoghue${}^{2}$}
\email[Email: ]{donoghue@physics.umass.edu}
\affiliation{${}^{1}$Department of Physics,
University of Toronto\\
Toronto, ON, M5S1A7, Canada\\
${}^{2}$Department of Physics,
University of Massachusetts\\
Amherst, MA  01003, USA\\
}

\begin{abstract}
We display several examples of how fields with different limiting velocities (the "speed of light") at a high energy scale can nevertheless have a common limiting velocity at low energies due to the effects of interactions. We evaluate the interplay of the velocities through the self-energy diagrams and use the renormalization group to evolve the system to low energy. The differences normally vanish only logarithmically, so that an exponentially large energy trajectory is required in order to satisfy experimental constraints. However, we also display a model in which the running is power-law, which could be more phenomenologically useful. The largest velocity difference should be in system with the weakest interaction, which suggests that the study of the speed of gravitational waves would be the most stringent test of this phenomenon.
\end{abstract}
\maketitle


\section{Introduction}
Many physical systems yield wave-like solutions which satisfy the wave equation with a speed of propagation $c_i$,
\begin{equation}
\left[ \frac{\partial^2}{\partial t^2} - c_i^2 \nabla^2 \right]\phi (\mathbf{x},t) =0
\end{equation}
which is also the massless Klein-Gordon equation. To leading order, the Lagrangian of any such field obeys a Lorentz-like symmetry of Lorentz transformations scaled with the limiting speed $c_i$, even if the underlying system does not have that invariance. However, if there are multiple fields, they will in general have different limiting velocities, and there will not be a global Lorentz symmetry. If the fundamental interactions are emergent phenomenon from an underlying theory without Lorentz invariance\cite{emergent, horava, Moffat}, we might expect that particles would display different limiting speeds.

In this paper we show how interactions between the fields can lead to a universal limiting velocity, i.e. the speed of light, at low energies. We calculate how the different fields influence each other's propagation velocity through the self-energy diagrams, and then use the renormalization group to scale the results to low energy. Using several examples we show that the condition of equal velocities is the low energy endpoint of renormalization group evolution\footnote{During the course of this work we found that this general approach has also been studied by S.-S. Lee\cite{Lee:2006if} in the context of emergent supersymmetry. There is also some overlap of our work with the study of Lifshitz type theories in Ref. \cite{Iengo:2009ix}. The renormalization group running that we describe is also related to the study of the renormalization of Lorentz-violating electrodynamics\cite{Kostelecky:2001jc}.  }. A heuristic explanation for this is that because fields can split into other types of fields, the propagation velocity of one field approaches that of the related other fields.

This result could be useful if the fields of the Standard Model are emergent from an underlying theory that is not Lorentz invariant. Of course, Lorentz invariance is conventionally taken as one of the foundational principles underlying all our fundamental interactions. However, the Weinberg-Witten \cite{weinbergwitten} theorem is usually interpreted as telling us that non-Abelian gauge bosons and gravitons cannot be emergent fields arising from any underlying Lorentz invariant non-gauge theory. All known examples\cite{emergent} satisfy this property. Therefore if the idea of emergent fields has any application in the fundamental interactions it appears required that Lorentz invariance is also emergent.

Our results show that a universal limiting velocity can be an emergent property in the low energy limit. However, in general the difference in velocity runs towards zero only logarithmically. This means that the underlying scale of emergence needs to be exponentially far away, making it difficult to test any feature of that theory which is power-suppressed. For example, we estimate that the scale where differences in the velocity are of order 10\% would be beyond $10^{10^{13}}$~GeV. Because of this feature we propose a model that produces much faster power-law running. The model involves a large number of fields which accelerate the running. Another consequence of the running speeds is that observable differences in the velocities would be greatest if the interactions are the weakest. This suggests that the measurement of the velocity of gravitational waves would be the most sensitive test of this aspect of emergence.

This paper has the following structure. In the next section we give some general comments on our procedure. Then in Sec. 3, 4, 5 and 6 we calculate the beta functions for Yukawa theories, electrodynamics, Yang-Mills and mixed theories respectively. All cases yield beta functions such that the limiting velocities run towards each other at low energy. In Sec 7, we analyze the general effects of logarithmic running and address the phenomenological constraints. Because of the difficulties posed by logarithmic running, we display a model with power-law running in Sec. 8. We close with a summary and discussion. Some of the more technical details are described in a pair of appendices.

\section{Setup}

We assume that different species of fermions, scalars and gauge fields emerge at some UV scale with different speeds of light. In condensed matter systems, phonons and magnons do not propagate at the same speed. Similarly, the same behavior is expected to carry on in an emergent theory of nature. In the absence of any form of interactions between particles, their speeds are expected to be frozen  as we run down to the IR.  However, these particles are interacting due to Yukawa and gauge forces. Hence, the total Lagrangian of such a  system will be given by the sum of kinetic and interaction terms with certain {\em bare} coefficients specified initially at the UV.  The parameter space of the system is spanned by the different speeds and interaction strengths. According to the principle of self-similarity and  Wilsonian renormalization, the same Lagrangian will continue to describe the system at different energy scales provided that we replace the bare parameters with the
 {\em renormalized} ones. This can be achieved by integrating out the high momentum modes as we run down from UV to IR. Quantum loops are sensitive to high momenta, and hence can be used to track the evolution of trajectories of the different speeds and interaction strengths in the parameter space. The evolution of these trajectories are encoded in the $\beta$ functions that are given by the Gell-Mann Low equations
\begin{equation}
\beta(g_i)\equiv \mu\frac{dg_i}{d\mu}=f{\{g_j\}}\,,
\end{equation}
where $\mu$ is the mass scale we introduce in dimensional regularization.

In theories with a universal limiting velocity, the Lorentz symmetry prevents the renormalization of the speed of light, and one can set $c=1$ as a definition of natural units. However, if different species carry different limiting velocities, $c_i$, then these parameters also get renormalized and must be treated in the same manner as coupling constants. They carry a scale dependence through the renormalization procedure, and also generate their own beta function. We exploit this property to study the running of the limiting velocities.

\begin{figure}[ht]
\leftline{
\includegraphics[width=.45\textwidth]{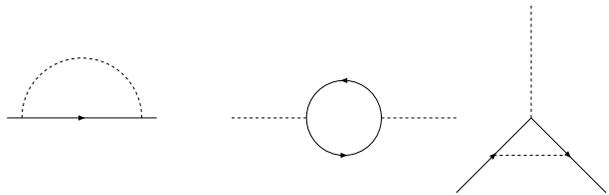}
}
\caption{The self-energy and vertex diagrams. Only self-energies will contribute to the running of the speeds, while the vertex is needed for the running of the coupling strength. }
\label{quantum loops}
\end{figure}

Throughout the paper we use dimensional regularization (dim-reg). The high energy part of the quantum loops can be isolated by retaining only the $1/\epsilon$ pieces that arise upon using dim-reg. Finally, we notice that our treatment is limited to one-loop corrections, and that the $\beta$ functions of the speeds require only self-energy corrections, while those of the couplings require vertex corrections as well (see Fig. \ref{quantum loops}).

\section{Yukawa interactions }

We consider a two-species system, namely scalars and fermions, having different speeds of light at the UV and coupled through Yukawa interaction. The Lagrangian density reads
\begin{eqnarray}
\nonumber
{\cal L}_r&=&i\bar \psi_r \gamma^0\partial_0\psi_r-ic_f\bar \psi_r\vec \gamma\cdot \vec \partial \psi_r+\frac{1}{2}\partial_0\phi_r\partial_0\phi_r\\
&&-\frac{c_b^2}{2}\vec \partial \phi_r\cdot \vec \partial \phi_r-g\phi_r\bar \psi_r \psi_r\,,
\end{eqnarray}
where the subscript $r$ denotes the renormalized values of the fields.
The momentum-space propagators for scalars and fermions are given by
\begin{eqnarray}
\nonumber
D_b\left(p^0,\vec p\right)&=&\frac{i}{(p^0)^2-c_b^2\vec p^2}\\
S_f\left( p^0,\vec p\right)&=&\frac{i}{p^0\gamma^0-c_f\vec p\cdot \vec \gamma}\,.
\end{eqnarray}
The self-energies of fermions and scalars are respectively
\begin{eqnarray}
\nonumber
-i\Sigma\left(p^0,\vec p \right)=&&\left(-ig\right)^2\int \frac{d^4 q}{\left(2\pi\right)^4}S_f\left(q^0,\vec q\right)\\
\label{fermionSE}
&&\times D_b\left(p^0-q^0,\vec p-\vec q\right)\,,
\end{eqnarray}
and
\begin{eqnarray}
\nonumber
i\Pi\left(p^0,\vec p\right)&=&-\left(-ig\right)^2\int \frac{d^4 q}{\left(2\pi\right)^4}\mbox{tr}\left[S_f\left(q^0,\vec q\right)\right.\\
\label{bosonSE}
&&\left.\times S_f\left(p^0+q^0,\vec q+\vec p\right) \right]\,.
\end{eqnarray}

In the following, we will be interested only in the divergent pieces of (\ref{fermionSE}) and (\ref{bosonSE}).  The integral  (\ref{bosonSE}) is trivial to perform upon using the substitutions $k^0=q^0/c_f$, $\vec k=\vec q$, $P^0=p^0/c_f$, and $\vec P=\vec p$. Then, one readily finds
\begin{equation}
i\Pi\left(p^0,\vec p\right)=\frac{ig^2}{8\pi^2c_f}\left[\frac{(p^0)^2}{c_f^2 }-\vec p^2  \right]\left(\frac{2}{\epsilon}+\mbox{finite}\right)\,.
\end{equation}
On the other hand, the integral (\ref{fermionSE}) is more involved and it needs a bit more attention.  Using the substitution $k^0=q^0/c_f$, and $\vec k=\vec q$ we find
\begin{eqnarray}
\nonumber
-i\Sigma&=&\frac{g^2}{c_b^2}\int \frac{d^4 k}{\left(2\pi\right)^4}\frac{\displaystyle{\not} k}{k^2}\frac{1}{\left(p^0/c_b-c_f k^0/c_b\right)^2-\left(\vec k -\vec p\right)^2}\\
\label{primary integrals}
&=&\frac{g^2}{c_b^2}\left[\gamma^0 I^0-\vec \gamma\cdot \vec I \right]\,,
\end{eqnarray}
where the integrals $I^0$ and $\vec I$ are given by (the details are in Appendix A)
\begin{equation}\label{I0}
I^0=\frac{ip^0}{\left(4\pi\right)^2}\frac{2c_b}{c_f^2(1+a)^2}\left(\frac{2}{\epsilon}+\mbox{finite}\right)\,,
\end{equation}
and
\begin{equation}\label{I1}
\vec I=\frac{i\vec p}{\left(4\pi\right)^2}\frac{2a(1+2a)}{3(1+a)^2}\left(\frac{2}{\epsilon}+\mbox{finite}\right)\,,
\end{equation}
where $a=c_b/c_f$.

Now, we move to the vertex correction which, to one-loop order, reads
\begin{eqnarray}
\nonumber
-igG&=&\left(-ig\right)^3\int \frac{d^4 q}{\left(2\pi\right)^4}S_f(p_2^0-q^0,\vec p_2-\vec q)\\
\label{vertex correction in Yukawa}
&&\times S_f(p_1^0-q^0,\vec p_1-\vec q)D_b(q^0,\vec q)\,.
\end{eqnarray}
Using the change of variables $q^0/c_f=K^0$, $\vec q=\vec K$, $p_2^0/c_f=P_2^0$, $\vec p_2=\vec P_2$, $p^0_{1}=P_1^0/c_f$, and $\vec p_1=\vec P_1$, and retaining only the divergent part of the integral we obtain (see Apendix A)
\begin{equation}\label{result for Yukawa vertex}
-igG=\frac{ig^3}{\left(4\pi\right)^2}\frac{2}{c_f^2c_b(1+a)}\left(\frac{2}{\epsilon}+\mbox{finite}\right)\,.
\end{equation}

At this point, the total Lagrangian including the one-loop effect is
\begin{eqnarray}
\nonumber
{\cal L}_0&=&{\cal L}_r+{\cal L}_{c}\\
\nonumber
&&+\frac{2ig^2}{\left(4\pi\right)^2(1+a)^2}\left[\frac{1}{c_bc_f^2}\bar \psi_r\partial_0\gamma^0\psi_r\right.\\
\nonumber
&&\left.-\frac{a(1+2a)}{3c_b^2}\bar \psi_r\vec\gamma\cdot \partial\psi_r \right]\left(\frac{2}{\epsilon}\right)\\
\nonumber
&&+\frac{g^2}{\left(4\pi\right)^2c_f}\left[\frac{1}{c_f^2}\partial_0\phi_r\partial_0\phi_r-\vec\partial\phi_r\cdot \vec\partial\phi_r \right]\left(\frac{2}{\epsilon} \right)\\
\label{full Yukawa}
&&+\frac{g^3}{\left(4\pi\right)^2c_bc_f^2}\frac{2}{1+a}\left(\frac{2}{\epsilon} \right)\phi_r\bar \psi_r\psi_r\,,
\end{eqnarray}
where ${\cal L}_c$ is the counter Lagrangian
\begin{eqnarray}
\nonumber
{\cal L}_c&=&i\delta_{Z_\psi}\bar \psi_r \partial_0\gamma^0\psi_r-i\delta_{Z_f}c_f\bar \psi_r \vec\gamma\cdot \vec\partial\psi_r\\
\nonumber
&&+\frac{\delta_{Z_\phi}}{2}\partial_0\phi_r\partial_0\phi_r-\frac{\delta_{Z_b}}{2}c_b^2\vec \partial \phi_r\cdot \vec\partial\phi_r-g\delta_{g}\phi_r\bar\psi_r\psi_r\,. \\
\end{eqnarray}
 At this point, we can read off the different $\delta$s that are required to absorb the infinities. Furthermore, we define the bare fields $\phi_0=Z_\phi^{1/2}\phi_r$, and $\psi_0=Z_\psi^{1/2}\psi_r$, bare speeds $c_{f
_0}$, and $c_{b_0}$ and bare coupling $g_0$ such that the Lagrangian density reads
\begin{eqnarray}
\nonumber
{\cal L}_0&=&i\bar \psi_0\partial_0\gamma^0\psi_0-c_{f0}\bar \psi_0\vec \gamma\cdot\vec \partial \psi_0+\frac{1}{2}\partial_0\phi_0\partial_0\phi_0\\
\label{bare Yukawa}
&&-c_{b0}^2\vec\partial \phi_0\cdot \vec\partial \phi_0-g_0\phi_0\bar\psi_0\psi_0\,.
\end{eqnarray}
Comparing (\ref{full Yukawa}) and (\ref{bare Yukawa}) we find
\begin{eqnarray}
\nonumber
c_{f_0}&=&c_fZ_\psi^{-1} Z_f\,,\\
\nonumber
c_{b_0}&=&c_bZ_\phi^{-1/2}Z_b^{1/2}\,,\\
\label{relation between the bare and the dressed in Yukawa}
g_0&=&gZ_gZ_\phi^{-1/2}Z_\psi^{-1}\mu^{\epsilon/2}\,,
\end{eqnarray}
where $Z=1+\delta$
\begin{eqnarray}
\nonumber
Z_\psi&=&1-\frac{2g^2}{\left(4\pi\right)^2c_b(c_f+c_b)^2}\left(\frac{2}{\epsilon}\right)\,,\\
\nonumber
Z_\phi&=&1-\frac{2g^2}{\left(4\pi\right)^2c_f^3}\left(\frac{2}{\epsilon}\right)\,,\\
\nonumber
Z_f&=&1-\frac{2g^2(c_f+2c_b)}{3\left(4\pi\right)^2c_fc_b(c_f+c_b)^2}\left(\frac{2}{\epsilon}\right)\,,\\
\nonumber
Z_b&=&1-\frac{2g^2}{\left(4\pi\right)^2c_fc_b^2}\left(\frac{2}{\epsilon}\right)\,,\\
\label{Z for Yukawa system}
Z_g&=&1+\frac{2g^2}{\left(4\pi\right)^2c_fc_b(c_f+c_b)}\left(\frac{2}{\epsilon}\right)\,.
\end{eqnarray}

To proceed, we regard all the renormalized quantities above as functions of the scale $\mu$ that occurs in dim-reg. Then, we differentiate the system in Eq. \ref{relation between the bare and the dressed in Yukawa} w.r.t $\mu$ and solve simultaneously for $\beta(g)$, $\beta(c_b)$ and $\beta(c_f)$ to find
\begin{eqnarray}
\nonumber
\beta(g)&=&\frac{g^3\left(3c_bc_f^2+2c_b^2c_f+c_b^3+4c_f^3\right)}{8\pi^2c_bc_f^3(c_f+c_b)^2}\,,\\
\nonumber
\beta(c_b)&=&\frac{g^2\left(c_b^2-c_f^2\right)}{8\pi^2c_bc_f^3}\,,\\
\label{the beta function of the simple yukawa system}
\beta(c_f)&=&\frac{g^2(c_f-c_b)}{6\pi^2c_b(c_f+c_b)^2}\,.
\end{eqnarray}
Notice that  the $\beta$ functions of $c_b$ and $c_f$ do not depend on the vertex correction $Z_g$.
Finally, we calculate the $\beta$ function of the ratio $a=c_b/c_f$ to find
\begin{eqnarray}
\nonumber
\beta(a)&=&\frac{\beta(c_b)}{c_f}-\frac{c_b}{c_f^2}\beta(c_f)\\
&=&\frac{g^2}{48\pi^2}\frac{(a-1)\left[8a+6(1+a)^3\right]}{c_bc_f^2(1+a)^2}\,,
\end{eqnarray}
from which we see that $c_b=c_f$ is an IR attractive line. We can also see that by studying the Jacobian $J= \partial\beta(c_i)/\partial c_j$, for $i,j=c_f,c_b$ at the fixed line $c_f=c_b$. The eigenvalues of $J$ are $\{0,7g^2/24\pi^2c_f^3 \}$. The positivity of the second value ensures that $c_b=c_f$ is an IR attractive fixed line.

We have seen the existence of an attractive IR fixed line corresponding to a common limiting speed. We will address more details about the running in Sec. 7.

\section{ Non-covariant Electrodynamics}

In this section, we study the RG flow of the limiting speeds of fermions and photons. The non-covariant and gauge invariant Lagrangian density reads
\begin{eqnarray}
\nonumber
{\cal L}_r&=&-\frac{1}{4}F_{r\,\mu\nu}F^{\,\mu\nu}_r+i\bar \psi_r \left(\partial_0+iec_gA_{0\,r}\right)\gamma^0\psi_r\\
&&-i\bar\psi_r\left(c_f\vec \partial +iec_f\vec A_r\right)\cdot\vec\gamma \psi_r\,,
\end{eqnarray}
where $F_{r\,\mu\nu}=\partial_\mu A_{r\,\nu}-\partial_\nu A_{r\,\mu}$, and $\partial_\mu=(\partial_0,c_g\vec \partial)$, and $c_g$ is the photon speed.
The photon propagator in the Feynman gauge is given by
\begin{equation}
D_{g\,\mu\nu}(k^0,\vec k)=\frac{-i\eta_{\mu\nu}}{\left(k_0\right)^2-c_g^2\vec k^2}\,.
\end{equation}

To find the photon and fermion self-energies, it proves easier to write the interaction Lagrangian in the form ${\cal L}_I=-ec_{\mu\nu}\bar \psi_r A^\mu_r\gamma^\nu \psi_r$, where $c_{\mu\nu}=\mbox{diag}(c_g,-c_f,-c_f,-c_f)$. Hence, the fermion self-energy is
\begin{eqnarray}
\nonumber
-i\Sigma(p^0,\vec p)&=&\left(-ie\right)^2c_{\beta\mu}c_{\alpha\nu}\gamma^\mu\int\frac{d^4 q}{\left(2\pi\right)^4} S_f(q^0,\vec q)\\
\nonumber
&&\times D_{g}^{\alpha\beta}(p^0-q^0,\vec p-\vec q)\gamma^\nu\,,\\
\nonumber
&=&-e^2\eta^{\alpha\beta}\frac{c_{\beta\mu}c_{\alpha\nu}}{c_g^2}\gamma^\mu\left[\gamma^0I^0-\vec \gamma\cdot\vec I\right]\gamma^\nu\,,\\\label{fermion self energy in general electrodynamics}
\end{eqnarray}
where $I_0$ and $I_1$ are given in (\ref{I0}) and (\ref{I1}) after replacing $c_b$ with $c_g$. While the photon self-energy is given by
\begin{eqnarray}
\nonumber
i\Pi_{\alpha\beta}\left(p^0,\vec p \right)&=&-(-ie)^2c_{\alpha\nu}c_{\beta\mu}\int \frac{d^4 q}{\left(2\pi\right)^4}\mbox{tr}\left[\gamma^\nu S_f\left(q^0,\vec q\right) \right.\\
&&\left. \gamma^\mu S_f\left(p^0+q^0,\vec p+\vec q\right) \right]\,.
\end{eqnarray}
Using the substitution $k_0=q^0/c_f$, $\vec k=\vec q$, $P_0=p^0/c_f$, and $\vec P=\vec p$ we can put $\Pi_{\alpha\beta}$ in a standard integral form. Hence,
\begin{eqnarray}
\nonumber
i\Pi_{\alpha\beta}\left(p^0,\vec p \right)&=&\frac{4i}{3\left(4\pi\right)^2}\frac{e^2c_{\alpha\nu}c_{\beta\mu}}{c_f}\left(P^\mu P^\nu-P^2\eta^{\mu\nu}\right)\\
\label{final form of photon self energy}
&&\times\left(\frac{2}{\epsilon}\right)\,,
\end{eqnarray}
where $P=(p^0/c_f,\vec p)$. Explicit calculations shows that ${\cal P}^\alpha\Pi_{\alpha\beta}=0$, where ${\cal P}^\alpha=\left(p^0,c_g\vec p\right)$, and hence $\Pi_{\alpha\beta}$ is gauge invariant as expected.

The counter Lagrangian reads
\begin{eqnarray}\label{counter lagrangian general electrodynamics}
\nonumber
{\cal L}_c&=&{\cal L}_{c\,\mbox{\scriptsize gauge}}+i\delta_{Z_\psi}\bar \psi_r \partial_0\gamma^0\psi_r-i\delta_{Z_f}c_f\bar \psi_r \vec\gamma\cdot \vec\partial\psi_r\,,\\
\end{eqnarray}
and ${\cal L}_{c\,\mbox{\scriptsize gauge}}$ is the counter term for the gauge sector. Then, from (\ref{counter lagrangian general electrodynamics}) and (\ref{fermion self energy in general electrodynamics}), and after using the properties of $\gamma$ matrices, we can immediately read $Z_\psi$ and $Z_f$
\begin{eqnarray}
\nonumber
Z_\psi&=&1-\frac{2e^2(3c_f^2-c_g^2)}{\left(4\pi\right)^2c_g(c_f+c_g)^2}\left(\frac{2}{\epsilon}\right)\,,\\
\nonumber
Z_f&=&1-\frac{2e^2(c_g^2+c_f^2)(2c_g+c_f)}{3\left(4\pi\right)^2c_fc_g(c_g+c_f)^2}\left(\frac{2}{\epsilon}\right)\,.
\end{eqnarray}

Now we come to the counter terms in the gauge sector. A general  counter term written in the momentum space takes the form
\begin{eqnarray}
\nonumber
{\cal L}_{c\,\mbox{\scriptsize gauge}}(p)&=&A_{0\,r}\delta_A\left[ \left(p^0\right)^2 -\eta^{00}\left(\left(p^0\right)^2 -c_g^2 \vec p^{\,2} \right)  \right]A_{0,r}\\
\nonumber
&&+A_{i\,r}\left[c_g^2\delta_{g_B}p^ip^j\right.\\
\nonumber
&&\quad\quad\quad\left.+\delta^{ij}\left(\delta_A \left(p^0\right)^2-c_g^2\delta_{g_B}\vec p^{\,2}\right) \right]A_{j\,r}\\
&&-2A_{i\,r}\delta_A c_g p^0p^iA_{0\,r}\,.
\end{eqnarray}
One can show that all the infinities in (\ref{final form of photon self energy}) can be absorbed using $\delta_A$ and $\delta_{g_B}$
\begin{eqnarray}
\nonumber
Z_{A}=1-\frac{4}{3\left(4\pi\right)^2}\frac{e^2}{c_f}\left(\frac{2}{\epsilon}\right)\,,\\
\label{Z for QED}
Z_{g_B}=1-\frac{4}{3\left(4\pi\right)^2}\frac{e^2c_f}{c_g^2}\left(\frac{2}{\epsilon}\right)\,.
\end{eqnarray}
where as usual $Z=1+\delta$. We write ${\cal L}_{c\,\mbox{\scriptsize gauge}}(p)$ in the compact form
\begin{eqnarray}
{\cal L}_{c\,\mbox{\scriptsize gauge}}(p)=A_{\mu\,r}M^{\mu\nu}A_{\nu\,r}\,,
\end{eqnarray}
with
\begin{eqnarray}
\nonumber
M^{00}&=&c_g^2\delta_A\vec p^{\,2}\,,\\
\nonumber
M^{0i}&=&-c_g\delta_Ap^0p^i\,,\\
\nonumber
M^{ij}&=&c_g^2\delta_{g_B}p^ip^j+\delta^{ij}\left(\delta_A\left(p^0\right)^2-c_g^2\delta_{g_B}\vec p^{\,2}\right)\,.\\\label{counter terms in the gauge sector of QED}
\end{eqnarray}
It is trivial to see that ${\cal P}_\alpha M^{\alpha\beta}=0$, and hence $M_{\alpha\beta}$ is gauge invariant.

Now, defining the bare fields $\psi_0=Z_\psi^{1/2}\psi_r$, $A_0^0=Z_{A}^1Z_{g_B}^{-1/2}A_r^0$, and $A_0^i=Z_A^{1/2}A_r^i$, and bare speeds $c_{f0}$, and $c_{g0}$ the Lagrangian density reads
\begin{eqnarray}
\nonumber
{\cal L}_{0}&=&-\frac{1}{4}{\cal F}_{0\,\mu\nu}{\cal F}_0^{\mu\nu}+i\bar\psi_0\partial_0\gamma^0\psi_0-ic_{f0}\bar\psi_0\vec \gamma\cdot \vec \partial \psi_0\,.\\
\end{eqnarray}
The bare gauge field Lagrangian in the momentum space is given by $A_{0\mu}M^{\mu\nu}_0A_{0\nu}$, and
\begin{eqnarray}
\nonumber
M_0^{00}&=&c_{g0}^2\vec p^{\,2}\,,\\
\nonumber
M_0^{0i}&=&-c_{g0}p^0p^i\,,\\
\nonumber
M_0^{ij}&=&=c_{g0}^2p^ip^j+\delta^{ij}\left(\left(p^0\right)^2-c_{g0}^2\vec p^{\,2}\right)\,.\\
\end{eqnarray}
The relations between the bare and renormalized speeds are
\begin{eqnarray}
\nonumber
c_{f0}&=&c_f Z_\psi^{-1}Z_f\,,\\
c_{g0}&=&c_g Z_{A}^{-1/2}Z_{g_B}^{1/2}\,,
\end{eqnarray}
from which we obtain
\begin{eqnarray}
\nonumber
\beta(c_g)&=&\frac{4e^2}{3\left(4\pi\right)^2}\frac{\left(c_g^2-c_f^2\right)}{c_fc_g}\,,\\
\beta(c_f)&=&\frac{8e^2}{3\left(4\pi\right)^2}\frac{\left(c_f-c_g\right)\left(4c_f^2+3c_fc_g+c_g^2\right)}{c_g\left(c_f+c_g\right)^2}\,.
\end{eqnarray}
These $\beta$ functions have the same structure as in the case of Yukawa interactions, and we immediately conclude that $c_f=c_g$ is an IR attractive line.

\section{Non-covariant Yang-Mills theories}

In this section we generalize the results of QED to the case of non-abelian gauge theories. We take the gauge group to be $SU(N)$, and the fermions in the fundamental representation
\begin{eqnarray}
\nonumber
{\cal L}_r&=&{\cal L}_{r\,\mbox{\scriptsize free}}\\
\nonumber
&&+gc_{\mu\nu}\bar \psi_r A^{a\mu}_{r}\gamma^\nu \psi_r t^a-gc_gf^{abc} \partial_\kappa A_{r\lambda}^aA_r^{\kappa b}A^{\lambda c}_r\\
&&-\frac{1}{4}g^2c_g^2f^{eab}f^{ecd}A_{r\kappa}^aA_{r\lambda}^bA_r^{c\kappa}A_r^{d\lambda}\,.
\end{eqnarray}
where $g$ is the coupling constant, $t^a$ are the group generators and $f^{abc}$ are the group structure constants.  ${\cal L}_{r\,\mbox{\scriptsize free}}$ is the free part of the Lagrangian
\begin{eqnarray}
\nonumber
{\cal L}_{r\,\mbox{\scriptsize free}}&=&-\frac{1}{4}F_{r\,\mu\nu}^aF^{\,a\mu\nu}_r+i\bar \psi_r\partial_0\gamma^0\psi_r\\
&&-ic_f\bar\psi_r\vec \partial \cdot\vec\gamma \psi_r\,,
\end{eqnarray}
where as in the case of QED  $F^a_{r\,\mu\nu}=\partial_\mu A^a_{r\,\nu}-\partial_\nu A^a_{r\,\mu}$, $\partial_\mu=(\partial_0,c_g\vec \partial)$, and $c_g$ is the gauge boson speed.

The fermion self-energy is identical to the case of QED, one just includes the quadratic Casimir operator in the fundamental representation $C_2(N)=(N^2-1)/2N$ into Eq. \ref{fermion self energy in general electrodynamics} to find
\begin{eqnarray}
\nonumber
Z_\psi&=&1-\frac{2C_2(N)g^2(3c_f^2-c_g^2)}{\left(4\pi\right)^2c_g(c_f+c_g)^2}\left(\frac{2}{\epsilon}\right)\,,\\
\nonumber
Z_f&=&1-\frac{2C_2(N)g^2(c_g^2+c_f^2)(2c_g+c_f)}{3\left(4\pi\right)^2c_fc_g(c_g+c_f)^2}\left(\frac{2}{\epsilon}\right)\,.
\end{eqnarray}

In calculating the gauge boson self-energy $\Pi_{\alpha\beta}^{ab}$ one encounters, in addition to the fermion loop, gauge boson and ghost loops
\begin{eqnarray}
\nonumber
i\Pi_{\alpha\beta}^{ab}&=&\left[-i\frac{5C_2(G) g^2}{3\left(4\pi\right)^2 c_g}\left({\cal P}_\alpha{\cal P}_\beta-{\cal P}^2\eta_{\alpha\beta}\right)\right.\\
\nonumber
&&+\left.i\frac{4 C(N)g^2c_{\alpha\nu}c_{\beta\mu}}{3c_f}\left(P^\mu P^\nu-\eta^{\mu\nu}P^2 \right)\right]\delta^{ab}\\
&&\quad \times \left(\frac{2}{\epsilon}\right)\,,
\end{eqnarray}
where $C_2(G)=N$, and $C(N)=1/2$ are group factors, ${\cal P}^\mu=\left(p^0,c_g\vec p\right)$ and $P^\mu=\left(p^0/c_f,\vec p \right)$. As we did in QED, the infinites can be absorbed into the counter term $A^a_{\mu\,r}M^{\mu\nu}A^a_{\nu\,r}$ where $M^{\mu\nu}$ are given in  Eq. \ref{counter terms in the gauge sector of QED}. Hence, we find
\begin{eqnarray}
\nonumber
Z_A&=&1+\left(-\frac{4 C(N) g^2}{3\left(4\pi\right)^2c_f}+\frac{5 C_2(G) g^2}{3\left(4\pi\right)^2c_g}\right)\left(\frac{2}{\epsilon}\right)\,,\\
\nonumber
Z_{g_B}&=&1+\left(-\frac{4 C(N) g^2 c_f}{3\left(4\pi\right)^2c_g^2}+\frac{5C_2(G)g^2}{3\left(4\pi\right)^2c_g}\right)\left(\frac{2}{\epsilon}\right)\,.\\
\end{eqnarray}
Gluon loops will not modify their own propagation speed, due to the Lorentz-like symmetry of that sector when considered in isolation. This is visible in the formulas above. Since $\beta(c_g)\propto (Z_{g_B}-Z_A)$, the gauge bosons and ghosts contributions  cancel in obtaining $\beta(c_g)$. Overall, the $\beta$ functions read
\begin{eqnarray}
\nonumber
\beta(c_g)&=&\frac{4C(N)g^2}{3\left(4\pi\right)^2}\frac{\left(c_g^2-c_f^2\right)}{c_fc_g}\,,\\
\nonumber
\beta(c_f)&=&\frac{8C_2(N)g^2}{3\left(4\pi\right)^2}\frac{\left(c_f-c_g\right)\left(4c_f^2+3c_fc_g+c_g^2\right)}{c_g\left(c_f+c_g\right)^2}\,,\\
\end{eqnarray}
which, apart from group factors, are identical to the QED case.

\section{Emergence of Lorentz symmetry in a mixed system}

The emergence of a universal Lorentz Symmetry in the above examples is intriguing to explore a more general setup consisting of multi-species and/or mixing between fermions, bosons and gauge fields. Before delving into the most general case we  derive a general formula that enables us to calculate the $\beta$ functions of such complex systems. This is done in Appendix B.

\subsection{Yukawa-Electrodynamics}

Now, let us consider the more general case of Yukawa-electrodynamics. In this theory a fermion couples to a scalar through Yukawa interaction, and minimally to a $U(1)$ gauge field. The scalar is neutral under the $U(1)$ field. We assume that the fermion, scalar and gauge field all have different speeds of light, $c_f$, $c_b$ and $c_g$ respectively. This is the simplest generalization of the above cases. The scalar and gauge field self-energies are identical to their expressions in Yukawa and QED sections, while the fermion self-energy is the sum of the contributions from the scalar and gauge field.  The calculations of the corresponding $Z$ renormalizations are very straight forward, and can be obtained directly from the previous two sections. Thus, $Z_\phi$, $Z_b$, $Z_A$, and $Z_{g_B}$ are given by their expressions in Eq. \ref{Z for Yukawa system} and  Eq. \ref{Z for QED} respectively, while
\begin{eqnarray}
\nonumber
Z_{\psi}&=&1+\left(-\frac{2e^2(3c_f^2-c_g^2)}{\left(4\pi\right)^2c_g(c_f+c_g)^2}-\frac{2g^2}{\left(4\pi\right)^2c_b(c_b+c_f)^2}\right)\\
\nonumber
&&\quad\times\left(\frac{2}{\epsilon}\right)\,,\\
\nonumber
Z_f&=&1+\left(-\frac{2e^2(c_g^2+c_f^2)(2c_g+c_f)}{3\left(4\pi\right)^2c_fc_g(c_g+c_f)^2}\right.\\
&&\left.-\frac{2g^2(c_f+2c_b)}{3\left(4\pi\right)^2c_fc_b(c_f+c_b)^2}\right)\left(\frac{2}{\epsilon}\right)\,,
\end{eqnarray}
The relations between the bare and renormalized quantities are given as before
\begin{eqnarray}
\nonumber
c_{f0}&=&c_fZ_\psi^{-1}Z_f\,,\quad c_{b0}=c_bZ_{\phi}^{-1/2}Z_b^{1/2}\,,\\
\nonumber
c_{g0}&=&c_gZ_{A}^{-1/2}Z_{g_B}^{1/2}\,.
\end{eqnarray}
In order to find the $\beta$ functions of $c_b$, $c_f$, and $c_g$ we use eq. (\ref{the grand formula}) to find
\begin{eqnarray}
\nonumber
\beta_{c_f}&=&c_f\left[g\frac{\partial}{\partial g}+e\frac{\partial}{\partial e} \right](\rho_f-\rho_\psi)\,,\\
\nonumber
\beta_{c_b}&=&-\frac{c_bg}{2}\frac{\partial}{\partial g}(\rho_\phi-\rho_b)\,,\\
\label{structure of beta for yukawa electrodynamics}
\beta_{c_g}&=&-\frac{c_ge}{2}\frac{\partial}{\partial e}(\rho_A-\rho_{g_B})\,,
\end{eqnarray}
where $Z=1+\rho(2/\epsilon)$.
Finally, the $\beta$ functions read
\begin{eqnarray}
\nonumber
\beta(c_f)&=&\frac{g^2(c_f-c_b)}{6\pi^2c_b(c_f+c_b)^2}\\
\nonumber
&&+\frac{8e^2(c_f-cg)(4c_f^2+3c_fc_g+c_g^2)}{3\left(4\pi\right)^2c_g(c_g+c_f)^2}\,,\\
\nonumber
\beta(c_b)&=&\frac{g^2(c_b^2-c_f^2)}{8\pi^2c_f^3c_b}\,,\\
\beta(c_g)&=&\frac{4e^2}{3\left(4\pi\right)^2}\frac{\left(c_g^2-c_f^2\right)}{c_fc_g}\,.
\end{eqnarray}
This is exactly expected: since only fermions can couple to both scalars and gauge fields, we find that the photon and scalars speeds of light are identical to those found before, while the fermion speed gets contributions from both Yukawa and gauge sectors.

We can see that  $c_f=c_b=c_g$ is an IR attractive fixed line by computing the eigenvalues of the Jacobian $J=\left(\partial \beta_i/\partial c_j\right)|_{c_f=c_b=c_g}$, where $i,j=c_f,c_b,c_g$,
\begin{eqnarray}
\nonumber
\{0, \frac{\left(7g^2+12e^2\right)\pm\sqrt{\left(7g^2+12e^2\right)^2-304g^2e^2}}{48\pi^2}>0 \}\,.\\
\end{eqnarray}
%

\subsection{The general case}

We consider $N_f$ fermions interacting with $N_b$ scalars or gauge bosons. Although we shall carry out the calculation in the case of Yukawa interactions, the abelian and non-abelian $\beta$-functions have the same structure as we pointed out above.

 The general Lagrangian density reads
\begin{eqnarray}
{\cal L}&=&i\bar \psi_a \gamma^0\partial_0\psi_a-ic_{f_a}\bar \psi _a\vec \gamma\cdot \vec\partial\psi_a+\frac{1}{2}\partial_0\phi_i\partial_0\phi_i\\
\nonumber
&&-\frac{1}{2}c_{b_i}^2\vec \partial \phi_i\cdot \vec \partial \phi_i-\bar \psi_a\left(u_{ab}^i+i\gamma^5v_{ab}^i \right)\psi_b\phi_i\,,
\end{eqnarray}
where summation over repeated indices is implied. Denoting $z_{ab}^i=u_{ab}+iv_{ab}^i$ and noticing that $\bf z^i\rightarrow \bf z^{i\dagger } $ as we move $\bf z^i$ across the vertex $\phi \bar \psi \psi$ \cite{Coleman:1973sx} we find
\begin{eqnarray}
\nonumber
Z_{\psi_{a}}&=&1-\frac{2}{\left(4\pi\right)^2}\sum_{c,i}\frac{z_{ac}^iz^{i*}_{ca}}{c_{b_i}(c_{f_c}+c_{b_i})^2}\left(\frac{2}{\epsilon}\right)\,,\\
\nonumber
Z_{f_{a}}&=&1-\frac{2}{3\left(4\pi\right)^2}\sum_{c,i}\frac{z_{ac}^iz_{ca}^{*i}\left(c_{f_c}+2c_{b_i}\right)}{c_{f_a}c_{b_i}(c_{f_c}+c_{b_i})^2}\left(\frac{2}{\epsilon}\right)\,,\\
\nonumber
Z_{\phi_i}&=&1-\frac{16}{\left(4\pi\right)^2}\sum_{a,b}\frac{z_{ab}^iz_{ba}^{*i}}{(c_{f_a}+c_{f_b})^3}\left(\frac{2}{\epsilon}\right)\,,\\
\nonumber
Z_{b_i}&=&1-\frac{8}{3\left(4\pi\right)^2}\sum_{a,b}\frac{z_{ab}^iz_{ba}^{*i}\left(c_{f_b}^2+4c_{f_a}c_{f_b}+c_{f_a}^2 \right)}{c_{b_i}^2(c_{f_a}+c_{f_b})^3}\\
&&\times \left(\frac{2}{\epsilon}\right)\,.
\end{eqnarray}
$Z_{f_a}$, $Z_{\psi_a}$ can be read directly from the Yukawa expressions in Eq. \ref{Z for Yukawa system}, while $Z_{\phi_i}$ and $Z_{b_i}$ are obtained using a series of integrals similar to those given in Appendix A.
Notice that quantum loops can also generate off-diagonal corrections $Z_{\psi_a\psi_b}$ if the couplings $z^i_{ab}$ contain off-diagonal components.  These corrections will induce kinetic mixing terms of the form $i\alpha_{ab}\bar \psi_a \partial^0\gamma^0\psi_b+i\beta_{ab}\bar \psi_a \partial^i\gamma^i\psi_b$.  In Lorentz invariant theories, where $\alpha_{ab}=\beta_{ab}$, we can always find basis where $\alpha_{ab}=\beta_{ab}$ are diagonal by performing $SO(N_f)$  rotations. However, in the present case, and since in general $\alpha_{ab}\ne\beta_{ab}$, we have the freedom to diagonalize either the time-time or the space-space components. Diagonalizing the time-time component, and hence working in a basis where we have canonical kinetic terms, will always leave space-space mixing terms. We assume that these terms are always small compared to the diagonal speeds, i.e. $\beta_{ab}/c_{f_a}<<1$ for all $a$, and we ignore their evolution in the following analysis. The same thing
 can also be said about kinetic mixing terms for bosons.

To be able to use the grand formula (\ref{the grand formula}) we define
\begin{eqnarray}
z_{ab}^i=g_{3iab}\,,\quad z_{ab}^{*i}=g_{\bar 3iab}\,,\quad c_{f_a}=g_{1a}\,,\quad c_{b_i}=g_{2i}\,,
\end{eqnarray}
and
\begin{eqnarray}
\nonumber
Z_{\phi_i}=Z_{1i}\,,\quad Z_{b_i}=Z_{2i}\,,\quad Z_{z^i_{ab}}=Z_{3iab}\,,\\
Z_{z^{* i}_{ab}}=Z_{\bar 3iab}\,, \quad Z_{f_a}=Z_{4a}\,,\quad Z_{\psi_a}=Z_{5a}\,.
\end{eqnarray}
Now, we write (\ref{the grand formula}) as
\begin{eqnarray}\label{modified grand formula}
\beta_{\mu M}=2g_{\mu M}\sum_{\nu N}n_{\nu N,\mu M}\sum_{\alpha O}p_{\alpha O}g_{\alpha O}\frac{\partial \rho_{\nu N}}{\partial g_{\alpha O}}\,,
\end{eqnarray}
where the Greek incices run from $1$ to $3$ and $\bar 3$, and the upper case lattin letters run over $a$ and $i$. The non zero values in (\ref{modified grand formula}) are $n_{1i',2i}=-\delta_{ii'}/2$, $n_{2i',2i}=\delta_{ii'}/2$, $n_{4a',1a}=\delta_{aa'}$,  $n_{5a',1a}=-\delta_{aa'}$, $p_{3,abi}=1/2$, and $p_{\bar3,abi}=1/2$. Hence, we obtain
\begin{eqnarray}
\nonumber
\beta(c_{b_i})&=&\frac{8}{3\left(4\pi\right)^2c_{b_i}}\\
\nonumber
&&\times\sum_{a,b}\frac{z_{ab}^iz_{ba}^{*i}\left[6c_{b_i}^2-\left(c_{f_a}^2+4c_{f_a}c_{f_b}+c_{f_b}^2\right) \right]}{(c_{f_a}+c_{f_b})^3}\,,\\
\beta(c_{f_a})&=&\frac{4}{3\left(4\pi\right)^2}\sum_{ib}\frac{z_{ab}^iz_{ba}^{*i}\left[3c_{f_a}-c_{f_b}-2c_{b_i}\right]}{c_{b_i}\left(c_{f_b}+c_{b_i}\right)^2}\,.
\end{eqnarray}

\begin{figure}[ht]
\leftline{
\includegraphics[width=.48\textwidth]{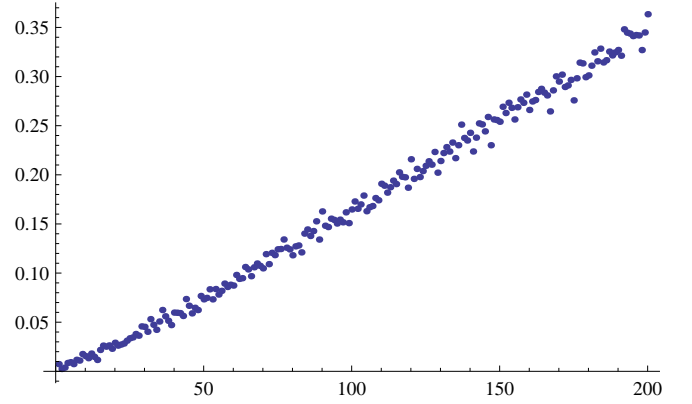}
}
\caption{ The smallest eigenvalue of $J$ (vertical) against the number of fermions $N_f$ (horizontal) taking $N_b=1$. The calculations are based on values of $z_{ab}^i$ between $0$ and $1$ generated randomly at each $N_f$. The plot shows that $\lambda \propto N_f$ for large $N_f$.  }
\label{eigenvalue vs Nf}
\end{figure}

Again, we find in this general setup that $c_{f_a}=c_{b_i}=c$ for all $ i\, \mbox{and}\, a$ is a fixed line. To study the nature of this line we perform a perturbation to the system about this line, i.e. we construct the Jacobian matrix at $c_{f_a}=c_{b_i}=c$. Defining $\Lambda_a^i=z^i_{aa}z^{*i}_{aa}$, and $\Gamma_a^i=\sum_{b\neq a}z_{ab}^iz_{ba}^{*i}$ we find
\begin{eqnarray}
J_{(N_f+N_b)\times (N_f+N_b)}=\left[\begin{array}{cc}
J^1_{N_f\times N_f}&J^2_{N_f\times N_b}\\
J^3_{N_b\times N_f}&J^4_{N_b\times N_b}
\end{array}\right]\,,
\end{eqnarray}
where
\begin{eqnarray}
\nonumber
J^1_{a,a}&=&\frac{\sum_{i}\left[2\Lambda_a^i+3\Gamma_a^i \right]}{3\left(4\pi\right)^2c^3}\,,\quad
J^1_{a,c}=-\frac{\sum_i z_{ac}^iz_{ca}^{*i}}{3\left(4\pi\right)^2c^3}\,,\\
\nonumber
J^2_{a,i}&=&\frac{-2\left[\Lambda_a^i+\Gamma_a^i\right] }{3\left(4\pi\right)^2c^3}\,,\quad
J^3_{i,a}=\frac{-4\left[\Lambda_a^i+\Gamma_a^i\right] }{\left(4\pi\right)^2c^3}\,,\\
J^4_{i,j}&=&\delta_{i,j}\frac{4\sum_{a}\left[\Lambda_a^i+\Gamma_a^i\right] }{\left(4\pi\right)^2c^3}\,.
\end{eqnarray}
Although we were not able to diagonalize $J$ analytically, numerical calculations show that we always have a spectrum of positive eigenvalues on the top of a zero mode. In figure (\ref{eigenvalue vs Nf}) we plot the smallest eigenvalue $\lambda$ of $J$, which governs the behavior of $c_{f_a}$ and $c_{b_i}$, against the number of fermions $N_f$ for fixed number of bosons $N_b$. We find that $\lambda \propto N_f$ for large $N_f$. Hence, the effective running increases with the number of species as expected.

\section{Implications of logarithmic running}

Let us consider how the differing speeds approach each other by studying the situation where the speeds are relatively close, but not identical, at some scale $\mu_*$. We treat this problem to first order in the speed difference. If we define the relative speed difference as
\begin{equation}
\eta = \frac{c_b}{c_f} -1\,,
\end{equation}
the generic beta function has the form
\begin{equation}
\beta (\eta) =\mu \frac{\partial \eta}{\partial \mu} =\frac{bg^2}{4\pi^2c^3} \eta + {\cal O}(\eta^2)\,,
\end{equation}
where $b$ is a constant of order unity and where we denote the common low energylimit as $c_f\approx c_b \approx c$.

If the coupling $g$ were to be treated as a constant, this running could be integrated to yield
\begin{equation}
\eta(\mu) = \eta_* \left(\frac{\mu}{\mu_*}\right)^{\frac{bg^2}{4\pi^2 c^3}}\,.
\end{equation}
The rescaling is power-law in form. If the coupling constant is small, the power-law exponent is also small and the running is slow. However, if the coupling is large (and constant) the running would be rapid with a power-law form, leading quickly to a universal speed of light.

However, the coupling itself also runs. For example, the Yukawa coupling beta function can be integrated to yield
\begin{equation}
\frac{g^2(\mu)}{4\pi c^3} = \frac{4\pi g_*^2}{5 \log(\frac{\Lambda^2}{\mu^2})}\,,
\end{equation}
with $\Lambda > \mu$.
While this coupling could be large at high energy it runs to smaller values at low energy\footnote{Clearly for the running coupling in Yang-Mills theory, the coupling is small at high energies and becomes large only at low energy. However the essential point is the same - that the coupling constant does not remain large at all energy scales.}. This produces a quite different form for the running of the relative speeds. The correct form for the running of $\eta$ is
\begin{equation}
\eta(\mu) = \eta_* \left[\frac{\log\left(\frac{\Lambda^2}{\mu_*^2}\right)}{\log\left(\frac{\Lambda^2}{\mu^2}\right)}\right]^{\frac{2b}{5}} =\eta_* \left[\frac{g^2({\mu})}{g^2(\mu_*)}\right]^{\frac{2b}{5}} \end{equation}\,.
This implies that the difference in the speeds runs only logarithmically towards each other.

There are tight constraints on the equality of the limiting velocities for the different particles. For direct measurement of the velocities, we can look at timing accomplished at high energy accelerators. For example at LEP, the electron beam travels at essentially the limiting velocity, since $E/m = \gamma \approx 10^5$. The timing of the accelerator relies on this limiting velocity being the speed of light. Because the timing of each bunch is recorded within $\pm 50$~ns over about 1000 revolutions in the 27~km accelerator\cite{Baribaud}, we estimate that this constrains $\eta \le 10^{-7}$ for electrons.

However, indirect constraints are more powerful, and these have been described by Altschul\cite{Altschul}. For $c_e>c$, energetic electrons traveling faster than the speed of light will radiate Cherenkov light, losing energy until they move at only the speed of light. This effect produces a maximum energy for subluminal motion, which is constrained by the observation of energetic electrons in astrophysics. For $c_e<c$, there is a constraint from the cutoff frequency in synchrotron emission. These constraints are more powerful than direct measurements because they bound factors of $\gamma_c = 1/\sqrt{1-c_e^2/c^2}\approx 1/\sqrt{\eta}$ rather than the linear bounds on $\eta$ from the velocity measurements. Altschul's bounds are $|\eta | \lesssim 10^{-14}$.

In order to achieve this close equality of the different speeds with logarithmic running, the running needs to occur over an exponentially large energy range. For example, even if we take $\eta_* \sim 10^{-1}$ and $\Lambda/\mu_* \sim 2$ (which barely allows perturbation theory to be used near the energy $\mu_*$), we would need $\log(\Lambda/m_e) \sim 10^{13}$, where we have generously used $m_e$ as the low energy scale. This clearly poses a problem for model building.

\section{Emergent Lorentz symmetry: toward Model building}

In any realistic model of emergence without an intrinsic Lorentz invariance, we do not expect the different species to emerge with the same limiting speed. In the above sections we showed that there is a potential mechanism to overcome this problem in a class of models whenever we run the renormalization group down to lower energies.  However, we found that the speeds of light are forced to run logarithmically along with the running coupling constants. This is a relatively slow running if we want to meet the stringent constraints on Lorentz violations without having to fine tune the speeds at the UV. In this section, we propose a way out of this situation.

In order to increase the effect of RG running there are two options. One is to keep the coupling constant large and unchanged with energy scale. Such a nearly conformal theory would convert logarithmic running into power-law running, as we saw in the last section. We also need the large coupling such that the exponent is large. Such theories are under active investigation \cite{walking} in the context of ``walking Technicolor'' where slowly running but strongly interacting theories are used to provide dynamical breaking of the Electroweak Theory while not producing excessive flavor changing processes. Should walking Technicolor theories prove successful, it would be quite interesting to tie those results with the idea of an emergent limiting velocity. The other option is if there are a very large number of fields of different scales, such that the running is increased by a large (and energy dependent) factor. We explore this option below.

We introduce a large number $N_{f}$ of hidden fermions in addition to the  Standard Model (SM) ones \cite{Dvali}. Moreover, we assume that all these fermions (hidden and SM) have the same origin, and hence all have the same initial speed of light $1+c_{f_*}$, with $|c_{f_*}|<<1$ , at some UV emergence scale $\mu_*$. As a warm up calculation, we assume that the fermions have a common initial charge $e_*$ under a single $U(1)$ gauge sector. The gauge photon emerges with some initial speed $1+c_{g_*}$, with $|c_{g_*}|<<1$ , that is different from the speed of fermions. At the UV scale the  fermions are taken to be massless and hence will participate in the running of the gauge coupling as well as photon and fermions speeds. As we run down our RG equations, some of the hidden fermions become massive and decouple from the RG equations. We model the dependence on the mass scale using a power law
\begin{eqnarray}
\label{power law running formulas}
N_{f}(\mu)&=&\Gamma_{f}\left(\frac{\mu}{M_{f}}\right)^{\alpha_{f}}\,,
\end{eqnarray}
where $\Gamma_{f}$ and $\alpha_{f}$ are positive constants, and $M_{f}$ is an IR mass scale.
\footnote{This exact behavior is also exhibited in models of large extra-dimensions where the Kaluza-Klein (KK) modes (from the 4D point of view) obey a power law as in Eq. \ref{power law running formulas} \cite{Masip:2000xy}. In this context, $M_f$ is the lowest KK mode $M_f\sim 1/L$, where $L$ is the size of the extra dimension, and $\alpha=d$ is the number of extra dimensions.}
 Since the fermions have a common initial speed and a common initial coupling strength, the evolution of the system can be modeled with a single $c_f$ and $e$ common to all fermions. Therefore, to one-loop order we have
\begin{eqnarray}
\nonumber
\mu\frac{de}{d\mu}&=&\frac{e^3}{12\pi^2}N_f(\mu)\,,\\
\nonumber
\mu\frac{dc_f}{d\mu}&=&\frac{e^2(c_f-c_g)}{3\pi^2}\,,\\
\label{RG phenomenology}
\mu\frac{dc_g}{d\mu}&=&\frac{e^2(c_g-c_f)}{6\pi^2}N_f(\mu)\,.
\end{eqnarray}
Integrating this system yields the running charge
\begin{eqnarray}\label{running of u(1) charge }
e^2(\mu)=\frac{e_*^2}{1+\frac{e_*^2\Gamma_f}{6\pi^2\alpha_f}\left[\left(\frac{\mu_*}{M_f}\right)^{\alpha_f}-\left(\frac{\mu}{M_f}\right)^{\alpha_f}\right]}\,,
\end{eqnarray}
and speeds
\begin{eqnarray}
\nonumber
\frac{c_g(\mu)-c_f(\mu)}{c_{g_*}-c_{f_*}}&=&\frac{e^2(\mu)}{e_*^2}\left[\left(\frac{e^2(\mu)}{e_*^2}\right)\left(\frac{\mu}{\mu_*}\right)^{\alpha_f} \right]^{\frac{e^2(0)}{3\pi^2\alpha_f}}\,\\
&\approx& \frac{e^2(\mu)}{e_*^2}\,.
\end{eqnarray}
Hence, we see that both $e(\mu)$ and $c_g(\mu)-c_f(\mu)$ experience power-law running with IR values given by
\begin{eqnarray}
\frac{c_g(0)-c_f(0)}{c_{g_*}-c_{f_*}} \approx\frac{e^2(0)}{e_*^2}\approx \frac{6\pi^2 \alpha_f}{e_*^2 \Gamma_f}\left(\frac{M}{\mu_*}\right)^{\alpha_f}\,.
\end{eqnarray}
Therefore, we can choose $\mu_*/M\sim 10^{14/\alpha_f}$ in order to meet the stringent requirement $\eta\sim 10^{-14}$. However, taking $e_*^2 \lesssim 1$, so that we can trust our perturbation theory, weakens the coupling strength to values $\sim 10^{-14}$. This is way below any interesting phenomenology.

\begin{figure}[ht]
\leftline{
\includegraphics[width=.48\textwidth]{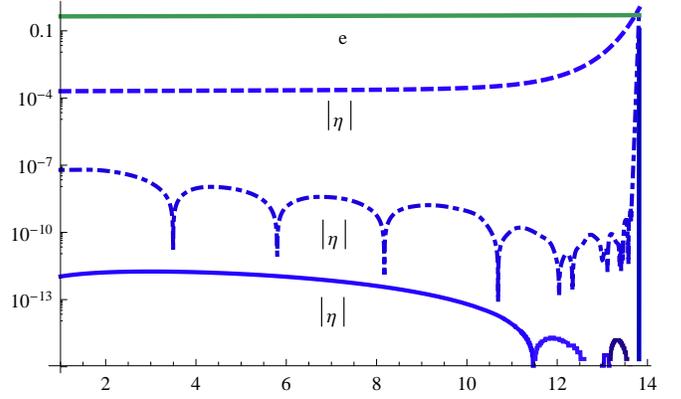}
}
\caption{Numerical simulation of the running of the charge $e$, and ratio $\eta=1-c_f/c_g$. The horizontal axis is in units of $\log (\mu/M_f)$. We take $M_{g}=M_{f}=1$, $\Gamma_g=10^{-3}$,$\Gamma_{f}=5$, $\alpha_f=10^{-2}$, $\mu_*/M_f=10^6$, and  $\alpha_g=1$, $1.2$, and $2$ for the dashed, dot-dashed and continuous lines respectively. We also use the initial conditions $c_{f_*}=0.6$, $c_{g_*}=0.3$, and $e_*=0.5$. The running of $e$ is logarithmic, while the running of $\eta$ is power-law. Very small values of $\eta$ are achieved in a very short interval of running as we increase $\alpha_g$, i.e. as we increase the number of gauge sectors. We also note that for $\alpha_g=2$ the value of $\eta$ is saturated by the error tolerance of the code. More powerful computations may give smaller values.}
\label{RG flow of speeds and coupling}
\end{figure}

In order to cure this problem, we introduce a large number of hidden $U(1)$ sectors in addition to the SM $U(1)_{hyp}$. We also assume that all these gauge sectors emerge with the same initial speed of light. In addition, we take all fermions to be charged under the different $U(1)$s with the same initial charge. As in the case of fermions, we assume that the hidden $U(1)$'s are massless at the UV scale, then they become massive and decouple as we run down the RG equations. Hence, the gauge sector obeys the scale-dependent relation
\begin{eqnarray}\label{power law running for gauge bosons}
N_{g}(\mu)&=&\Gamma_{g}\left(\frac{\mu}{M_{g}}\right)^{\alpha_{g}}\,.
\end{eqnarray}
Under these assumptions the second Eq. in (\ref{RG phenomenology}) is replaced by
\begin{eqnarray}
\mu\frac{dc_f}{d\mu}&=&\frac{e^2(c_f-c_g)}{3\pi^2}N_g(\mu)\,.
\end{eqnarray}
The solution of $e(\mu)$ is still given by Eq. \ref{running of u(1) charge }, while that of $c_g(\mu)-c_f(\mu)$ can be expressed in terms of the hypergeometric function. Instead, we numerically integrate our system setting  appropriate values of the parameters $\alpha_f$, $\alpha_g$, $\Gamma_f$, and $\Gamma_g$. From  FIG. \ref{RG flow of speeds and coupling} we see that introducing many copies of $U(1)$s achieves power-law suppression of $\eta$ in a very short interval. Since the many $U(1)$s do not intrude the evolution of $e$, we can still get reasonable coupling strength in the IR.

To understand the choice of parameters used in the simulation in FIG. \ref{RG flow of speeds and coupling}, it is instructive to calculate the total number of fermions and gauge fields as seen in the UV. Using Eqs.  \ref{power law running formulas} and \ref{power law running for gauge bosons}, and the numerical coefficients given in FIG. \ref{RG flow of speeds and coupling} (take $\alpha_g=2$) we find $N_g\approx 10^9$, and $N_f\approx 6$, which explains the above findings. Because the RG flow  of $e$ is enhanced only by a few number of fermionic species, the coupling constant runs only logarithmically. While the huge number of gauge fields that participate in the RG equation of speeds force the running of $c_g-c_f$ to be extremely fast.

Because the freezing of the speeds sets in almost immediately, we can try to replace the power law numbers in Eqs. \ref{power law running formulas} and \ref{power law running for gauge bosons} by constant numbers. The idea is that we just need to have $N_{g}>>N_{f}>>1$ for a short interval until the speeds freeze to their desired values. Then, we can immediately integrate the RG equations to find
\begin{eqnarray}
\nonumber
e^2(\mu)&=&\frac{e_*^2}{1+\frac{N_f e_*^2}{6\pi^2}\log\left(\frac{\mu_*}{\mu}\right)}\,,\\
\frac{c_g(\mu)-c_f(\mu)}{c_{g_*}-c_{f_*}}&\cong&\left(\frac{e^2(\mu)}{e_*^2}\right)^{2N_g/N_f}\,.
\end{eqnarray}
Now if we take $e_*\approx 1$, $\mu_*/\mu_{IR}\approx 10^2$, $Nf \approx 120$, and $N_g\approx 900$ we find that $e_{IR}^2/4\pi\approx 1/129$, and $\eta<10^{-14}$. Moreover, assuming that the masses of the hidden sectors are larger than $\mu_{IR}$, these masses decouple below $\mu_{IR}$ and drop from the RG equations. Therefore, one needs only a constant large number of copies, $N_{g}\,, N_{f}>>1$ to accomplish the emergence of an IR Lorentz-invariant fixed point in a short interval of running. Moreover choosing the ratio $N_{g}/N_{f}>>1$, we can meet the stringent constraints on the parameter $\eta$.  This opens up the possibility that many copies of hidden sectors may suppress Lorentz-violating effects already present at the TeV scale.

\section{Discussion}

Achieving a universal speed of light is a challenge for theories which do not postulate an fundamental Lorentz symmetry.
This problem is visible in known emergence models\cite{emergent} and also in Ho\v{r}ava-Lifshitz gravity\cite{horava}. For emergent gauge fields, the Weinberg-Witten theorem\cite{weinbergwitten} suggests that this will be a continual challenge as a Lorentz-noninvariant initial theory may be required.

We have shown through several examples that a common limiting velocity can be emergent at low energies even if the original high-energy theory involves fields satisfying the wave equation with different velocities. There is a heuristic rationale for this in that since fields can transform into each other through interactions, the endpoint where all the fields travel in unison is preferred. The renormalization group treatment indeed produces this outcome.

Because the running is only logarithmic for simple systems, it would take an exponentially large amount of running in order that the limiting velocities be
close enough to agree with experiment. We addressed the phenomenological constraints in Sec. 7. However, power law running is also possible if the coupling is large and constant, or if there are a very large number of interacting degrees of freedom.  We have reported on a model with this latter property.

It is important to note that not all forms of Lorentz-violation disappear at low energies. The renormalization of a general parameterization of Lorentz-violation of QED has been studied in Ref. \cite{Kostelecky:2001jc}, and some operators that grow at low energy are found. A well behaved emergent theory must avoid those operators.

The running of the limiting velocities only happens due to the interactions that couple one type of particle to another. This implies that the running will be weakest if the coupling is weak. At low energies the gravitational coupling is by far the weakest of all the fundamental forces. This implies that the most plausible velocity difference would be that of gravity. While there have been some claims that the speed of gravity has been indirectly measured\cite{wrong}, the consensus appears to be that there is no experimental constraint on the speed of gravity\cite{will}. However indirectly there is a stringent limit at the $10^{-15}$ level on the difference of the speeds of gravity and that of light from gravitational Cherenkov radiation  \cite{Moore:2001bv} which is valid if the speed of gravity is less than that of light. Future experiment with gravitational wave detectors provide the best opportunity to measure or constrain the difference if the speed of gravity is greater than that of light.

\section*{Acknowledgements}

M.A. would like to thank Erich Poppitz for stimulating discussions and suggestions,  and Bob Holdom and Amanda Peet for useful conversations. We thank Lorenzo Sorbo for bringing Ref. \cite{Moore:2001bv} to our attention. The work of M.A. is supported  by NSERC Discovery Grant of Canada.
The work of J.D. has been supported in part by the NSF grant PHY - 0855119, and in part by the Foundational Questions Institute.

\appendix
\renewcommand{\theequation}{A\arabic{equation}}
  \setcounter{equation}{0}  
  \section*{Appendix A: Useful Integrals}

In this appendix we work out the details of the Integrals $I^0$ and $\vec I$ appearing in Eq. \ref{primary integrals}. These integrals are given by
\begin{eqnarray}
\nonumber
I^0&=&\int\frac{dk^0}{2\pi}k^0\int \frac{d\vec k}{\left(2\pi\right)^3}\frac{1}{(k^0)^2-\vec k^2}\\
&&\times\frac{1}{\left(p^0/c_b-c_f k^0/c_b\right)^2-\left(\vec k -\vec p\right)^2}\,,
\end{eqnarray}
and
\begin{eqnarray}
\nonumber
\vec I&=&\int\frac{dk^0}{2\pi}\int \frac{d\vec k}{\left(2\pi\right)^3}\frac{\vec k}{(k^0)^2-\vec k^2}\\
&&\times\frac{1}{\left(p^0/c_b-c_f k^0/c_b\right)^2-\left(\vec k -\vec p\right)^2}\,.
\end{eqnarray}
To perform the integral $I^0$, we first use the Feynman trick to find
\begin{equation}
I^0=\int\frac{dk^0}{2\pi}k^0\int \frac{d\vec k}{\left(2\pi\right)^3}\int _{0}^1 dx \frac{1}{\left(\vec k^2-\Delta^2\right)^2}\,,
\end{equation}
where $\Delta^2=-x(1-x)\vec p^2+x(k^0)^2+(1-x)(p^0/c_b-c_fk^0/c_b)^2$. Next, we interchange the integrals $dx$ and $d\vec k$ and perform the integral over $d\vec k$ to find
\begin{equation}
I^0=-\frac{i}{\left(4\pi\right)^{3/2}}\Gamma\left(\frac{1}{2}\right)\int \frac{dk^0}{2\pi}k^0\int_0^1 dx\frac{1}{\sqrt{\Delta^2}}\,.
\end{equation}
Further, we exchange the integrals $dx$ and $dk^0$, and rearrange the integrands to find
\begin{eqnarray}
\nonumber
I^0&=&-\frac{i}{\left(4\pi\right)^{3/2}}\Gamma\left(\frac{1}{2}\right)\int_0^1 dx\frac{1}{\sqrt{x(1-c_f^2/c_b^2)+c_f^2/c_b^2}}\\
&&\times \int \frac{dk^0}{2\pi}\frac{k^0}{\sqrt{k_0^2+2k_0R_0-M^2}}\,,
\end{eqnarray}
where

\begin{eqnarray}
\nonumber
R_0=-\frac{(1-x)p^0c_f/c_b^2}{x(1-c_f^2/c_b^2)+c_f^2/c_b^2}\,,\\
M^2=\frac{-x(1-x)\vec p^2+(1-x)(p^0)^2/c_b^2}{x(1-c_f^2/c_b^2)+c_f^2/c_b^2}\,.
\end{eqnarray}
Then, we perform the integral over $dk^0$, after analytically continuing from $1$ to $d=1-\epsilon$ dimensions, to obtain
\begin{eqnarray}
\nonumber
I^0&=&\frac{i}{\left(4\pi\right)^2}\frac{c_f}{c_b^2}p^0\int_0^1 dx \frac{(1-x)^{1-\epsilon/2}\left(-1\right)^{-\epsilon/2}}{\left[x(1-c_f^2/c_b^2)+c_f^2/c_b^2 \right]^{3/2-\epsilon}}\\
\nonumber
&&\times\frac{\Gamma(\epsilon/2)}{\pi^{\epsilon/2}\left[x(p^0)^2/c_b^2-x\left(x+(1-x)c_f^2/c_b^2\right)\right]^{\epsilon/2}}\,.\\
\end{eqnarray}
Finally, we find
\begin{equation}
I^0=\frac{ip^0}{\left(4\pi\right)^2}\frac{2c_b}{c_f^2(1+a)^2}\left(\frac{2}{\epsilon}+\mbox{finite}\right)\,,
\end{equation}
where $a=c_b/c_f$. Similarly, we can show
\begin{equation}
\vec I=\frac{i\vec p}{\left(4\pi\right)^2}\frac{2a(1+2a)}{3(1+a)^2}\left(\frac{2}{\epsilon}+\mbox{finite}\right)\,.
\end{equation}

The vertex correction in Eq. \ref{vertex correction in Yukawa} results in the integral

\begin{eqnarray}
\nonumber
-igG&=&\frac{g^3}{c_fc_b^2}\int \frac{dK^0}{2\pi}\int \frac{d\vec K}{\left(2\pi\right)^3}\frac{(K^0)^2-\vec K^2}{c_f^2(K^0)^2/c_b^2-\vec K^2}\\
\nonumber
&&\times \frac{1}{\left[\left(P_1^0-K^0\right)^2-\left(\vec P_1-\vec K\right)^2\right]}\\
&&\times \frac{1}{\left[\left(P_2^0-K^0\right)^2-\left(\vec P_2-\vec K\right)^2\right]}\,.
\end{eqnarray}
Next, we use The Feynman trick to get
\begin{eqnarray}
\nonumber
-igG&=&\frac{2g^3}{c_fc_b^2}\int \frac{dK^0}{2\pi}\int \frac{d\vec K}{\left(2\pi\right)^3}\left(\vec K^2-(K^0)^2\right)\\
&&\times\int_0^1dx\int_0^{1-x}dy\frac{1}{\left[\vec K^2-\Delta^2 \right]^3}\,,
\end{eqnarray}
where
\begin{eqnarray}
\nonumber
\Delta^2&=&\left[1+y\left(-1+c_f^2/c_b^2\right)\right](K^0)^2\\
\nonumber
&&-2\left[(1-x-y)P_1^0+P_2^0x \right]K^0\\
\nonumber
&&+\left[(1-x-y)\vec P_1+x\vec P_2 \right]^2\\
&&+(1-x-y) P_1^2+x P_2^2\,.
\end{eqnarray}
Then, proceeding as we did before, we finally obtain the result in Eq.\ref{result for Yukawa vertex}.

\appendix
\renewcommand{\theequation}{B\arabic{equation}}
  \setcounter{equation}{0}  
  \section*{Appendix B: A general Setup to calculate the $\beta$ functions}

We assume that the parameter space is spanned by  $g_i$, $i,j=1,2,..C$ couplings (these could be coupling strengths as well as speeds). Quantum loops will generate $Z_m$, $m,l=1,2,...D$ corrections to the original Lagrangian, and we restrict our treatment to one loop order. In general, we may write
\begin{eqnarray}
\label{general expression for bare coupling}
g_{i0}=g_i(\mu)\prod_{m=1}^D Z_m^{n_{m,i}}(\mu)\mu^{\epsilon p_i}\,.
\end{eqnarray}
Taking the derivative of of (\ref{general expression for bare coupling}) w.r.t. $\mu$ we obtain
\begin{eqnarray}
\nonumber
&&g_i'(\mu)\prod_{m=1}^D Z_m^{n_{m,i}}(\mu)\mu^{\epsilon p_i}\\
\nonumber
&&+g_i(\mu)\sum_{l=1}^D n_{l,i}Z_l'(\mu)\prod_{m\neq l}^DZ_m^{n_{m,i}}(\mu)\mu^{\epsilon p_i}\\
\label{next step in general expression}
&&+g_i(\mu)\epsilon p_i\prod_{m=1}^D Z_m^{n_{m,i}}(\mu)\mu^{\epsilon p_i-1}=0\,.
\end{eqnarray}
Writing $Z_m(\mu)=1+\rho_m(\mu)\frac{2}{\epsilon}$ we find $Z_l'(\mu)=\frac{2}{\epsilon}\sum_{j=1}^{C}\frac{\partial \rho_l}{\partial g_j}g_j'(\mu)$. Also, using the definition $\beta_i(\mu)\equiv\mu\frac{\partial g_i(\mu)}{\partial \mu}$, eq. (\ref{next step in general expression}) becomes
\begin{eqnarray}
\nonumber
&&\beta_i(\mu)\prod_{m=1}^D\left(1+\frac{2}{\epsilon}\rho_m(\mu) \right)^{n_{m,i}}\\
\nonumber
&&+g_i(\mu)\sum_{l=1}^D\sum_{j=1}^C\frac{2}{\epsilon}n_{l,i}\frac{\partial\rho_l}{\partial g_j}\beta_j(\mu)\prod_{m\neq l}^DZ_{m}^{n_{m,i}}(\mu)\\
\label{third step general expression}
&&+g_i(\mu)p_i\epsilon \prod_m^D\left(1+\frac{2}{\epsilon}\rho_m(\mu)\right)^{n_{m,i}}=0\,.
\end{eqnarray}
Since we are only interested in one-loop corrections, we can ignore all terms ${\cal O}\left(1/\epsilon^2\right)$. Hence, eq. (\ref{third step general expression}) reads
\begin{eqnarray}
\nonumber
&&\beta_i(\mu)\left[1+\frac{2}{\epsilon}\sum_{m=1}^D\rho_m(\mu) n_{m,i}  \right]\\
\nonumber
&&+\frac{2}{\epsilon}g_i(\mu)\sum_{l=1}^{D}\sum_{j=1}^{C}n_{l,i}\beta_j(\mu)\frac{\partial \rho_l}{\partial g_j}\\
\label{fourth step general expression}
&&+g_i(\mu)p_i\epsilon\left[1+\frac{2}{\epsilon}\sum_{m=1}^D\rho_m(\mu)n_{m,i}\right]=0\,.
\end{eqnarray}
Now, we can rewrite eq. (\ref{fourth step general expression}) in the following simple  expression
\begin{eqnarray}
\nonumber
&&\beta_i(\mu)\left[1+\frac{2}{\epsilon}{\cal A}_i(\mu)\right]+\frac{2}{\epsilon}\sum_{j\neq i}^C{\cal C}_{ij}(\mu)\beta_j(\mu)\\
\label{simple expression general beta}
&&=-g_i(\mu)p_i\left[\epsilon+2\sum_{m=1}^D\rho_m(\mu)n_{m,i}\right]\,,
\end{eqnarray}
where
\begin{eqnarray}
\nonumber
{\cal A}_i(\mu)&=&\sum_{m=1}^D\left[\rho_m(\mu)+g_i(\mu)\frac{\partial \rho_m}{\partial g_i}\right]n_{m,i}\,,\\
{\cal C}_{ij}(\mu)&=&g_i(\mu)\sum_{m=1}^D\frac{\partial \rho_m}{\partial g_j}n_{m,i}\,.
\end{eqnarray}
Eq. (\ref{simple expression general beta}) can also be written in the matrix form
\begin{equation}
\label{matrix form for beta}
\stackrel{\leftrightarrow}{\cal M}(\mu)\stackrel{\rightarrow}\beta(\mu)=-\stackrel{\rightarrow}{gp}\left[\epsilon +2\sum_{m=1}^D\rho_m n_{m,i}\right]\,,
\end{equation}
where
\begin{eqnarray}
\stackrel{\leftrightarrow}{\cal M}(\mu)=\left(  \begin{array}{cccc}
1+\frac{2}{\epsilon}{\cal A}_1 & \frac{2}{\epsilon} {\cal C}_{12} &.... & \frac{2}{\epsilon} {\cal C}_{1C}\\
\frac{2}{\epsilon} {\cal C}_{21} & 1+\frac{2}{\epsilon}{\cal A}_2 &.... & \frac{2}{\epsilon} {\cal C}_{2C}\\
....&....&....&....\\
\frac{2}{\epsilon} {\cal C}_{C1}& \frac{2}{\epsilon} {\cal C}_{C2}&....& 1+\frac{2}{\epsilon}{\cal A}_C
\end{array}\right)\,.
\end{eqnarray}
The inverse of $\stackrel{\leftrightarrow}{\cal M}(\mu)$ is given by
\begin{eqnarray}
\nonumber
\stackrel{\leftrightarrow}{\cal M}^{-1}(\mu)&=&\left(  \begin{array}{cccc}
1-\frac{2}{\epsilon}{\cal A}_1 & -\frac{2}{\epsilon} {\cal C}_{12} &.... & -\frac{2}{\epsilon} {\cal C}_{1C}\\
-\frac{2}{\epsilon} {\cal C}_{21} & 1-\frac{2}{\epsilon}{\cal A}_2 &.... & -\frac{2}{\epsilon} {\cal C}_{2C}\\
....&....&....&....\\
-\frac{2}{\epsilon} {\cal C}_{C1}& -\frac{2}{\epsilon} {\cal C}_{C2}&....& 1-\frac{2}{\epsilon}{\cal A}_C
\end{array}\right)\\
&&+{\cal O}\left(\frac{1}{\epsilon^2}\right)\,.
\end{eqnarray}
Hence, solving for $\stackrel{\rightarrow}\beta$ from (\ref{matrix form for beta}) we obtain
\begin{eqnarray}
\beta_i=2g_ip_i\left({\cal A}_i-\sum_{m=1}^D\rho_m n_{m,i}\right)+2\sum_{j\neq i}^{C}{\cal C}_{ij}g_jp_j\,.
\end{eqnarray}
Finally, we rearrange the terms to find
\begin{equation}\label{the grand formula}
\beta_i=2g_i\sum_{m=1}^Dn_{m,i}\sum_{j=1}^{C}p_jg_j\frac{\partial \rho_m}{\partial g_j}\,.
\end{equation}

\end{document}